\documentclass[aps,prd,superscriptaddress,showpacs,showkeys,nofootinbib,nobibnotes,twocolumn,preprintnumbers]{revtex4}

\usepackage{amsmath}
\usepackage{amssymb}
\usepackage{graphicx}
\usepackage{color}
\usepackage{curves}
\usepackage{bm}  
\usepackage[normalem]{ulem}

\newcommand{\del}{\partial}

\begin{document}

\preprint{ADP-12-09/T776}

\title{Origin of $\Delta I=1/2$ Rule for Kaon Decays: QCD Infrared 
Fixed Point} 
\author{R.~J.~Crewther}
\email{rcrewthe@physics.adelaide.edu.au}
\affiliation{CSSM and ARC Centre of Excellence for Particle Physics at 
the Tera-scale, School of Chemistry and Physics, University of Adelaide, 
Adelaide SA 5005 Australia}
\author{Lewis~C.~Tunstall}
\email{lewis.tunstall@adelaide.edu.au}
\affiliation{CSSM and ARC Centre of Excellence for Particle Physics at 
the Tera-scale, School of Chemistry and Physics, University of Adelaide, 
Adelaide SA 5005 Australia}
\affiliation{Berkeley Center for Theoretical Physics, Department of Physics, 
University of California, Berkeley, CA 94720, U.S.A.}

\begin{abstract}
We replace ordinary chiral $SU(3)_L\times SU(3)_R$ 
perturbation theory $\chi$PT$_3$ by a new theory 
$\chi$PT$_\sigma$ based on a low-energy expansion about 
an infrared fixed point $\alpha^{}_\mathrm{IR}$ for 3-flavor QCD.
At $\alpha^{}_\mathrm{IR}$, the quark condensate 
$\langle \bar{q}q\rangle_{\mathrm{vac}} \not= 0$ induces \emph{nine} 
Nambu-Goldstone bosons: $\pi, K, \eta$ and a $0^{++}$ QCD dilaton 
$\sigma$. Physically, $\sigma$ appears as the $f_{0}(500)$ resonance,
a pole at a complex mass with real part $\lesssim m_K$. The 
$\Delta I=1/2$ rule for nonleptonic $K$-decays is then a 
\emph{consequence} of $\chi$PT$_\sigma$, with a $K_S\sigma$ coupling 
fixed by data for $K_{S}^{0} \to \gamma\gamma$ and 
$\gamma\gamma\rightarrow\pi\pi$. We estimate $R_\mathrm{IR} \approx 5$ 
for the nonperturbative Drell-Yan ratio 
$R = \sigma(e^{+}e^{-}\rightarrow\mathrm{hadrons})/ 
   \sigma(e^{+}e^{-}\rightarrow\mu^{+}\mu^{-})$ at $\alpha^{}_\mathrm{IR}$.
\end{abstract}

\pacs{12.38.Aw, 13.25.Es, 11.30.Na, 12.39.Fe}

\keywords{Nonperturbative QCD, Infrared fixed point, Dilaton, 
Chiral lagrangians, Nonleptonic kaon decays}

\maketitle

\begin{figure}[t]
\center\includegraphics[scale=1]{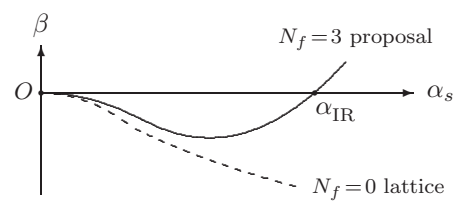}
\caption{The solid line shows a 3-flavor $\beta$ function (or 
better, a QCD version \cite{Grun82} of the Gell-Mann--Low $\Psi$ 
function) with an infrared fixed point $\alpha^{}_{\mathrm{IR}}$. 
Theoretical interest in $\alpha^{}_{\mathrm{IR}}$ \cite{Alk01} is
considerable but contradictory about $N_f$ dependence. The lattice 
result \cite{Lusc94} for $N_f = 0$ (no quarks) is that $\beta$ 
remains negative and becomes linear at large $\alpha_s$ 
(dashed line).}  
\label{fig:beta}
\end{figure}

The precise determination of the mass and width of the $f_0(500)$ resonance 
\cite{Cap06,Kam11, PDG} prompts us to revisit an old idea \cite{RJC70,Ell70}
that the chiral condensate $\langle \bar{q}q \rangle_{\mathrm{vac}} \not= 0$ 
may also be a condensate for scale transformations in the chiral 
$SU(3)_L\times SU(3)_R$ limit. This can occur in QCD if the heavy quarks 
$t,b,c$ are first decoupled and then the strong coupling%
\footnotemark[1]\addtocounter{footnote}{1}
\footnotetext[1]{We have $[D_\mu\,,\,D_\nu] 
= ig G^a_{\mu\nu}T^a$ where $D_\mu$ is the covariant derivative, $\{T^a\}$
generate the gauge group, $\alpha_s = g^2/4\pi$ is the strong 
coupling, and $\beta = \mu\del\alpha_{s}/\del\mu$ and 
$\delta = -\mu\del\ln m_q/\del\mu$ refer to a mass-independent
renormalization scheme with scale $\mu$.}%
$\alpha_s$ of the resulting 3-flavor theory runs nonperturbatively to a fixed
point $\alpha^{}_{\mathrm{IR}}$ in the infrared limit (Fig.\ \ref{fig:beta}).
At that point, $\beta(\alpha^{}_{\mathrm{IR}})$ vanishes, so the gluonic
term in the strong trace anomaly \cite{Mink76}
\begin{equation}\label{eqn:anomaly}
\theta^\mu_\mu
 =\frac{\beta(\alpha_{s})}{4\alpha_{s}} G^a_{\mu\nu}G^{a\mu\nu}
         + \bigl(1 - \delta(\alpha_{s})\bigr)\sum_{q=u,d,s} m_{q}\bar{q}q
\end{equation}
is absent, which implies
\begin{align}  
\left.\theta^\mu_\mu\right|_{\alpha_s = \alpha_{\mathrm{IR}}}
 &= \bigl(1 - \delta(\alpha_{\mathrm{IR}})\bigr)
    (m_u\bar{u}u + m_d\bar{d}d + m_s\bar{s}s) \nonumber \\
 &\to 0\ , \ SU(3)_L\times SU(3)_R \mbox{ limit}
\label{scale}\end{align}
and hence a $0^{++}$ QCD dilaton%
\footnotemark[2]\addtocounter{footnote}{1}
\footnotetext[2]{We reserve the term \emph{dilaton} and notation $\sigma$ 
for a Nambu-Goldstone boson due to \emph{exact scale invariance} 
in some limit. We are \emph{not} talking about the $\sigma$-model, scalar
gluonium \cite{glue}, or ``walking gauge theories'' \cite{Appel10} where
$\beta$ is small but never zero.}%
$\sigma$ due to quark condensation.%
\footnote{In field and string theory, it is often stated that Green's 
functions are manifestly conformal invariant for $\beta = 0$. That 
depends on the unlikely assumption that no scale condensates are 
present; otherwise, conformal invariance becomes manifest only if 
\emph{all} 4-momenta are space-like and large.}  
The obvious candidate for this state is the $f_0(500)$, which arises from 
a pole on the second sheet at a complex mass with typical value~\cite{Cap06}
\begin{equation}
m_{f_0} = 441-i\,272 \mbox{ MeV}
\label{f_0}\end{equation} 
and surprisingly small errors \cite{Mink10}.  In all estimates of this type, 
the real part of $m_{f_0}$ is less than $m_K$. 

\begin{figure}[b]
\center\includegraphics[scale=.7]{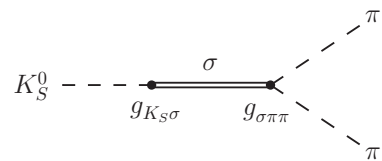}
\caption{Role of a QCD dilaton $\sigma$ in $K^0_S\to\pi\pi$ with couplings 
$g^{}_{K_S\sigma}$ and $g_{\sigma\pi\pi}$ derived from the effective theory 
$\chi$PT$_\sigma$.} \label{fig:k_sig_pipi}
\end{figure}

We begin by setting up chiral-scale perturbation theory 
$\chi$PT$_\sigma$ for amplitudes expanded in $\alpha_s$ 
about $\alpha^{}_{\mathrm{IR}}$.  Its Lagrangian summarises 
soft-$\{\pi, K, \eta, \sigma\}$ meson theorems for approximate 
chiral $SU(3)_L\times SU(3)_R$ and scale symmetry, with results 
for strong interactions similar to those found
originally \cite{RJC70,Ell70}. Effective weak operators are then added
to simulate nonleptonic $K$ decays.  The main result is a simple
explanation of the $\Delta I =1/2$ rule for kaon decays:  in 
\emph{leading order} of $\chi$PT$_\sigma$, there is a dilaton pole 
diagram (Fig.\ \ref{fig:k_sig_pipi}) which accounts for the dominant
$I=0$ amplitude.

It is well known that dispersive analyses which include $f_0(500)$ pole 
amplitudes produce excellent fits to data for $K \to \pi\pi$ 
\cite{Pol02,Ynd07,Trof12} and $\gamma\gamma \to \pi\pi$ 
\cite{Pen06,Oll08} as well as $\pi\pi \to \pi\pi$. The problem is that, 
in the context of conventional chiral $SU(3)_L\times SU(3)_R$ perturbation 
theory $\chi$PT$_3$, these corrections are \emph{far too large}.%
\footnote{Similarly $\chi$PT$_3$ fails for 
$K_L \to \pi^0\gamma\gamma$ \cite{Kam94}.} They cannot contribute to 
leading terms ${\cal A}_\mathrm{LO}$ in $\chi$PT$_3$ expansions
\begin{equation} 
{\cal A} = \bigl\{{\cal A}_\mathrm{LO} + {\cal A}_\mathrm{NLO} 
             + \ldots\bigr\}_{\chi\mathrm{PT}_3} 
\label{chiral}\end{equation}
because $f_0$ does not belong to the $\chi$PT$_3$ Goldstone sector 
$\{\pi,K,\eta\}$. How can this be reconciled with the success
\cite{Gasser85} of $\chi$PT$_3$ elsewhere%
\footnote{Except non-leptonic hyperon decays: either $\chi$PT for baryons 
or the weak sector of the Standard Model will have to be amended.}
where next to leading order terms ${\cal A}_\mathrm{NLO}$ are corrections 
$\lesssim$ 30\%?  The discrepancy is extreme for $K \to \pi\pi$: an $f_0$ 
pole in ${\cal A}_\mathrm{NLO}$ producing the factor of 22 enhancement 
of the $I=0$ amplitude would be $\approx$ 70 times the expected 30\% 
correction.
 
Our solution to this problem is to replace $\chi$PT$_3$ with  $\chi$PT$_\sigma$. 
We include $f_0 = \sigma$ in the Goldstone sector $\{\pi,K,\eta,\sigma\}$
and identify scale invariance (Eq.~(\ref{scale})) as the symmetry most 
likely to permit this. The idea works because $\chi$PT$_\sigma$ 
includes $f_0$ pole amplitudes in its \emph{leading order} terms without 
upsetting successful leading order $\chi$PT$_3$ predictions for amplitudes 
which do not involve the $0^{++}$ channel; that is because the $\chi$PT$_3$ 
Lagrangian equals the $\sigma \to 0$ limit of the $\chi$PT$_\sigma$ Lagrangian.  
In next to leading order, new chiral loop diagrams involving $\sigma$ need 
to be checked.

Notice that $\sigma$ becomes a Goldstone boson \emph{only} if all three
quarks $u,d,s$ become massless for $\alpha_s \to \alpha^{}_\mathrm{IR}$.

In the limit $m_{u,d} \to 0$ with $m_s \not= 0$, we use \emph{standard} 
chiral $SU(2)_L \times SU(2)_R$ perturbation theory $\chi$PT$_2$, 
where momenta $i\del$ are $O(m_\pi)$ and the Goldstone sector is 
$\{\pi^+, \pi^0, \pi^-\}$. We do \emph{not} include a dilaton in 
$\chi$PT$_2$: $f_0$ belongs to the non-Goldstone sector, so it retains most 
of its mass and width in that limit. $\chi$PT$_2$ is not sensitive 
to the behavior of $\beta$ because of the relatively large term 
$m_s\bar{s}s$ in Eq.~(\ref{eqn:anomaly}) for $\theta^\mu_\mu$.

Consider strong interactions at low energies 
$\alpha_s \lesssim \alpha^{}_\mathrm{IR}$ within the physical region
\begin{equation} 
0 < \alpha_s < \alpha^{}_{\mathrm{IR}} \,.
\label{phys}\end{equation}
The gluonic trace anomaly is represented by a term $\mathcal{L}_\mathrm{anom}$
in an effective chiral-scale Lagrangian
\begin{equation} 
\mathcal{L}\bigl[\sigma,U,U^\dagger\bigr] =\ :\mathcal{L}^{d=4}_\mathrm{inv} 
 + \mathcal{L}^{d>4}_\mathrm{anom} + \mathcal{L}^{d<4}_\mathrm{mass}:
\label{Lagr}\end{equation}
constructed from a chiral invariant QCD dilaton field 
$\sigma$ and the usual chiral $SU(3)$ field $U = U(\pi,K,\eta)$. 
Both $\mathcal{L}_\mathrm{inv}$ and $\mathcal{L}_\mathrm{anom}$ 
are $SU(3)_L \times SU(3)_R$ invariant, 
while $\mathcal{L}_\mathrm{mass}$ belongs to the representation 
$(\mathbf{3},\bar{\mathbf{3}})\oplus(\bar{\mathbf{3}},\mathbf{3})$ 
associated with the $\pi,K,\eta$ (mass)$^2$ matrix $M$.  The operator
dimensions of $\mathcal{L}_\mathrm{inv}$ and $\mathcal{L}_\mathrm{mass}$ 
satisfy $d_\mathrm{inv} = 4$ and $1 \leqslant d_\mathrm{mass} < 4$, with 
\begin{equation} 
d_\mathrm{mass} = 3 + \delta\bigl(\alpha^{}_\mathrm{IR}\bigr)
\label{dim-mass}\end{equation} 
as a result of expanding in $\alpha_s$ about $\alpha_\mathrm{IR}$.
The dimension of $\mathcal{L}_\mathrm{anom}$ is found by noting that 
the gluonic anomaly corresponds to the $\beta\del/\del\alpha_s$ 
term in the Callan-Symanzik equation
\begin{equation}
\mu\frac{\del\mathcal{A}}{\del\mu} + \beta(\alpha_s)
   \frac{\del\mathcal{A}}{\del\alpha_s}
=  \bigl(\delta(\alpha_{s}) - 1\bigr)\sum_q m_q
     \frac{\del\mathcal{A}}{\del m_q}
\end{equation}
for renormalization-group invariant QCD amplitudes $\mathcal{A}$.
Taking $\del/\del\alpha_s$, we find
\begin{equation}
\Bigl\{\mu\frac{\del\ }{\del\mu} + \beta\frac{\del\ }{\del\alpha_s} 
+ \beta'(\alpha_s)\Bigr\}\frac{\del{\cal A}}{\del\alpha_s} 
=  \sum_q m_q\frac{\del^2(\delta - 1){\cal A}}{\del m_q\del\alpha_s}
\end{equation}
so for $\alpha_s \lesssim \alpha^{}_\mathrm{IR}$, ${\cal L}_\mathrm{anom}$ 
has a positive anomalous dimension equal to the slope of $\beta$ at 
the fixed point (Fig.~\ref{fig:beta}):
\begin{equation}
d_\mathrm{anom} = 4 + \beta'\bigl(\alpha^{}_\mathrm{IR}\bigr) > 4\,.
\label{dim-anom}\end{equation}
As $\alpha_s \to \alpha^{}_\mathrm{IR}$, the gluonic anomaly vanishes, so 
for consistency, we must suppose that terms in ${\cal L}_\mathrm{anom}$ 
contain derivatives $O(\del\del) = O(M)$ or have $O(M)$ coefficients. The 
result is a chiral-scale perturbation expansion $\chi$PT$_\sigma$ about 
$\alpha^{}_\mathrm{IR}$ with QCD dilaton mass $m_\sigma = O(m_K)$.

Note that QCD in the limit (\ref{scale}) resembles the physical theory in 
the resonance region, but differs completely at high energies because it 
lacks asymptotic freedom: instead, Green's functions scale with 
nonperturbative anomalous dimensions. All particles except $\pi,K,\eta$ 
and $\sigma$ remain massive. Strong gluon fields set the scale of the 
condensate $\langle \bar{q}q \rangle_{\mathrm{vac}}$, which then sets the 
scale for massive particles and resonances except (possibly) glueballs. 


An explicit formula for the $\chi$PT$_\sigma$ Lagrangian (\ref{Lagr}) can
be found by applying the method of Ellis \cite{Ell70,Ell71}. Let $F_\sigma$ 
be the coupling of $\sigma$ to the vacuum via the energy momentum tensor 
$\theta_{\mu\nu}$, improved \cite{CCJ70} when spin-0 fields are present:
\begin{equation} 
\langle\sigma(q)|\theta_{\mu\nu}|\mathrm{vac}\rangle
= (F_\sigma/3)\bigl( q_\mu q_\nu - g_{\mu\nu}q^2\bigr) \,.
\label{b}\end{equation}
The dilaton field is given a scaling property $\sigma \to \sigma$ 
+ \{constant\} such that $e^{\sigma/F_\sigma}$ has dimension $1$.  
Then the dimensions of chiral Lagrangian operators such as 
\begin{equation}
\mathcal{K}\bigl[U,U^\dagger\bigr] 
= \tfrac{1}{4}F_{\pi}^{2}\mathrm{Tr}(\partial_{\mu} U\partial^{\mu}U^{\dagger})
\end{equation}
and the dilaton operator 
$\mathcal{K}_\sigma = \frac{1}{2}\partial_{\mu}\sigma\partial^{\mu}\sigma$
can be adjusted by powers 
of $e^{\sigma/F_\sigma}$ to form terms in $\mathcal{L}$. In leading order,
\begin{align}
&\mathcal{L}^{d=4}_\mathrm{inv,\,LO}
 = \bigl\{c_{1}\mathcal{K} + c_{2}\mathcal{K}_\sigma 
     + c_{3}e^{2\sigma/F_{\sigma}}\bigr\}e^{2\sigma/F_{\sigma}} \,,
\notag \\[1mm]
&\mathcal{L}^{d>4}_\mathrm{anom,\,LO} \notag \\
&= \bigl\{(1-c_{1})\mathcal{K} + (1-c_{2})\mathcal{K}_\sigma
      + c_4 e^{2\sigma/F_{\sigma}}\bigr\}e^{(2+\beta')\sigma/F_{\sigma}} \,,
\notag \\[1mm]
&\mathcal{L}^{d<4}_\mathrm{mass,\,LO} 
 = \mathrm{Tr}(MU^{\dagger}+UM^{\dagger})e^{(3+\delta)\sigma/F_{\sigma}} \,,
\label{Lstr}
\end{align}
where $\beta'$ and $\delta$ are the anomalous dimensions 
$\beta'(\alpha^{}_\mathrm{IR})$ and $\delta(\alpha^{}_\mathrm{IR})$ 
of Eqs.\ (\ref{dim-anom}) and (\ref{dim-mass}).  The constants $c_{1}$ 
and $c_{2}$ are not fixed by general arguments, while $c_3$ and $c_4$
depend on how the field $\sigma$ is chosen. For the vacuum to be stable 
in the $\sigma$ direction at $\sigma = 0$, terms linear in $\sigma$
must cancel:
\begin{align}
4c_3 + (4+\beta')c_4 &= - (3+\delta)\bigl\langle\mathrm{Tr}
        (MU^{\dagger}+UM^{\dagger})\bigr\rangle_{\mathrm{vac}} \notag \\
 &= - (3+\delta)F_\pi^2\bigl(m_K^2 + \tfrac{1}{2}m_\pi^2\bigr)\,.
\end{align}
Because of our requirement ${\cal L}_\mathrm{anom} = O(\del^2,M)$,  both 
$c_3$ and $c_4$ are $O(M)$. 

The critical exponent $\beta'$ normalises the gluonic term 
in the trace of the effective energy-momentum tensor:
\begin{equation}
\left.\theta^\mu_\mu\right|_\mathrm{eff} 
   =\ :\beta'\mathcal{L}^{d>4}_\mathrm{anom} 
      + (\delta - 1)\mathcal{L}^{d<4}_\mathrm{mass}: \,. \label{eff-tr}
\end{equation}

In leading order, $\cal L$ gives formulas for the $\sigma\pi\pi$ coupling 
\begin{equation}
\mathcal{L}_{\sigma\pi\pi}
 = \bigl\{\bigl(2+(1-c_1)\beta'\bigr)|\del\bm{\pi}|^2 
    - (3 + \delta)m_\pi^2|\bm{\pi}|^2\bigr\}\sigma/(2F_\sigma) \label{Lsigpi}
\end{equation}
and dilaton mass $m_\sigma$
\begin{equation}
m_\sigma^2 F_\sigma^2
= F_\pi^2\bigl(m_K^2 + \tfrac{1}{2}m_\pi^2\bigr)(3 + \delta)(1 - \delta)
  - \beta'(4 + \beta')c_4
\label{mass}\end{equation}
which resemble pre-QCD results \cite{Ell70,RJC70,Ell71,Klein71} but have
extra gluonic terms proportional to $\beta'$. We assume that the unknown 
coefficient $2+(1-c_1)\beta'$ in Eq.~(\ref{Lsigpi}) does not vanish 
accidentally. That preserves the key feature of the original work, that 
$\mathcal{L}_{\sigma\pi\pi}$ is mostly \emph{derivative}: 
for soft $\pi\pi$ scattering (energies $\sim m_\pi$), the dilaton pole
amplitude is negligible because the $\sigma\pi\pi$ vertex is 
$O(m_\pi^2)$, while the  $\sigma\pi\pi$ vertex for an on-shell dilaton
\begin{equation}
g_{\sigma\pi\pi} = -\bigl(2+(1-c_1)\beta'\bigr)m_\sigma^2/(2F_\sigma) 
+ O(m_\pi^2)
\label{on-shell}
\end{equation}
is $O(m_\sigma^2)$, consistent with $\sigma$ being the broad resonance 
$f_0(500)$.

Comparisons with data require an estimate of $F_\sigma$, most simply from
$NN$ scattering and the dilaton relation
\begin{equation}
-F_\sigma g_{\sigma NN} \approx M_N\,.
\end{equation}
The data imply \cite{CC08} a mean value $g_{\sigma NN} \sim 9$ and 
hence $F_{\sigma} \sim -100\,\mathrm{MeV}$ but with an uncertainty
which is either model dependent or very large ($\approx 70\%$).  That 
accounts for the large uncertainty in 
\begin{equation}
 1\tfrac{1}{2} \lessapprox |2 + (1-c_{1})\beta'| \lessapprox 6
\end{equation}
when we compare Eq.~(\ref{on-shell}) with 
$|g_{\sigma\pi\pi}| = 3.31^{+0.35}_{-0.15}\,\mathrm{GeV}$ \cite{Kam11}
and $m_{\sigma} \approx 441\,\mathrm{MeV}$.  

The convergence of our chiral-scale expansion can be tested by adding 
$\sigma$-loop diagrams to the standard analysis \cite{Gasser85,ManGeo} 
for $\chi$PT$_3$. These involve the (as yet) undetermined constants
$\beta',\delta,c_{1\ldots 4}$: for example, corrections to $g_{\sigma\pi\pi}$
involve the $\sigma\sigma\sigma$ and $\sigma\sigma\pi\pi$ vertices derived
from Eq.~(\ref{Lstr}). However, when we apply the dimensional arguments 
of Manohar and Georgi \cite{ManGeo} to our scheme, we find that there 
are \emph{two} $\chi$PT$_\sigma$ scales $\chi_\pi = 4\pi F_\pi$ and 
$\chi_\sigma = 4\pi F_\sigma$, which are numerically similar ($F_\sigma 
\sim  F_\pi$). The following points should be noted:
\begin{enumerate}
 \item The $\alpha_s \lessapprox \alpha_{IR}$ scales $\chi_{\pi,\sigma} 
       \approx$ 1 GeV must not be confused with the scale 
       $\Lambda_{\mathrm{QCD}} \approx$ 200 MeV for ultraviolet expansions 
       $\alpha_s \gtrapprox 0$ (asymptotic freedom).
 \item The small value of $F_\sigma \ll \chi_{\pi,\sigma}$ implies a $\sigma$ 
       width 
\begin{equation}
      \Gamma_{\sigma\pi\pi} \approx \frac{|g_{\sigma\pi\pi}|^2}{16\pi m_\sigma} 
      \sim \frac{m_\sigma^3}{16\pi F_\sigma^2} \sim  250 \mbox{ MeV} 
\end{equation}
       which is numerically misleading: $\Gamma_{\sigma\pi\pi}$ is 
       $O(m_\sigma^3)$ and hence \emph{non-leading} relative to the mass 
       $m_\sigma$. So tree diagrams produce the leading order%
\footnote{ Beyond leading order, and in degenerate cases like the 
       $K_{L}$--$K_{S}$ mass difference, methods used to estimate 
       corrections at the $Z^0$ peak \cite{ZOphys} and the $\rho$ 
       resonance \cite{Scherer} may be necessary.}
       of $\chi$PT$_\sigma$, as in $\chi$PT$_2$ and $\chi$PT$_3$. 
 \item The technique used to obtain Eq.~(\ref{Lstr}) from  $\chi$PT$_3$ also
       works for $O(\del^4, M\del^2, M^2)$ terms in $\cal L$ (and in 
       ${\cal L}_{\mathrm{weak}}$ below). For example, a term 
       $\sim (\mbox{Tr}\del U\del U^\dagger)^2$ appears in 
       ${\cal L}^{d=4}_{\mathrm{inv, NLO}}$ without $\sigma$-field dependence, 
       but within ${\cal L}^{d>4}_{\mathrm{anom, NLO}}$, it becomes%
       \footnote{To the extent allowed by data for soft-$\sigma$ 
       amplitudes, $O(\del^4)$ coefficients can be predicted by 
       saturation by non-Goldstone resonances --- like 
       $\chi$PT$_3$ \cite{Eck89}, but with $f_0$ {\it excluded}.}
\begin{equation}
\{\mbox{coefficient}\}\bigl(\mbox{Tr}\del U\del U^\dagger\bigr)^2
                e^{\beta'\sigma/F_\sigma}.
\end{equation}
\end{enumerate}

The most important feature of $\chi\mathrm{PT}_{\sigma}$ is that it
explains the $\Delta I=1/2$ rule for $K\rightarrow\pi\pi$ decays. 

In the leading order of standard $\chi$PT$_3$, the effective weak Lagrangian
\begin{equation}
\left.\mathcal{L}_{\mathrm{weak}}\right|_{\sigma=0} 
= g_{8}Q_{8} + g_{27}Q_{27} + Q_{M} + \mathrm{h.c.}\
\label{usual}\end{equation}
contains an octet operator \cite{Cro67}
\begin{equation}
Q_{8} = \mathcal{J}_{13}\mathcal{J}_{21} - \mathcal{J}_{23}\mathcal{J}_{11}
\ , \quad
\mathcal{J}_{ij} = (U\partial_{\mu}U^{\dagger})_{ij}
\end{equation}
the $U$-spin triplet component \cite{Gaill74,RJC86} of a \textbf{27} operator
\begin{equation}
Q_{27} = \mathcal{J}_{13}\mathcal{J}_{21} 
           + \tfrac{3}{2} \mathcal{J}_{23}\mathcal{J}_{11}
\end{equation}
and a weak mass operator \cite{Bern85}
\begin{equation}
Q_M = \mathrm{Tr} (\lambda_6 - i\lambda_7)
      \bigl(g_MMU^\dagger + \bar{g}_MUM^\dagger\bigr) \,.
\end{equation}
Although $Q_M$ has isospin 1/2, it cannot be used to solve the 
$\Delta I = 1/2$ puzzle if dilatons are absent: when $Q_M$ is added
to the strong mass term $\left.{\cal L}_\mathrm{mass}\right|_{\sigma = 0}$, 
it can be removed by a chiral rotation which aligns the vacuum 
\cite{RJC86} such that $\langle U \rangle_\mathrm{vac} = I$ and 
$M$ = real diagonal. Therefore the conclusion that $|g_8/g_{27}|$ 
is unreasonably large ($\approx$ 22) is not avoided.

In $\chi$PT$_\sigma$, the outcome is entirely different. The weak mass
operator's dimension $(3+\delta_w)$ is not the same as the dimension
$(3+\delta)$ of ${\cal L}_\mathrm{mass}$, so the $\sigma$ dependence of 
$Q_M e^{(3+\delta_w)/F_\sigma}$ cannot be eliminated by a chiral rotation.
Instead, after aligning the vacuum, we find
\begin{align}
\mathcal{L}^{\mathrm{align}}_{\mathrm{weak}}
 &= Q_{8}\sum_n g_{8n}e^{d_{8n}\sigma/F_\sigma} 
  + g_{27}Q_{27}e^{d_{27}\sigma/F_\sigma}   \nonumber \\
 &\ + Q_{M}\bigl\{e^{(3+\delta_w)\sigma/F_\sigma} - e^{(3+\delta)\sigma/F_\sigma}\bigr\} 
  + \mathrm{h.c.}\,,
\end{align}
noting that $Q_8$ represents quark-gluon operators with differing
dimensions at $\alpha_\mathrm{IR}$. As a result, there is a residual 
interaction $\mathcal{L}_{K_S\sigma} = g^{}_{K_S\sigma}K_{S}^{0}\sigma$
which mixes $K_{S}$ and $\sigma$ in \emph{leading order}
\begin{equation}
g^{}_{K_S\sigma} =  (\delta_w - \delta)\mathrm{Re}\{(2m^2_K - m^2_\pi)\bar{g}_M
                    - m^2_\pi g_M\}F_{\pi}/2F_{\sigma} 
\end{equation}
and produces the $\Delta I = 1/2$ amplitude $A_{\sigma\textrm{-pole}}$ of
Fig.~\ref{fig:k_sig_pipi}.
\begin{figure}[t]
\center\includegraphics[scale=.43]{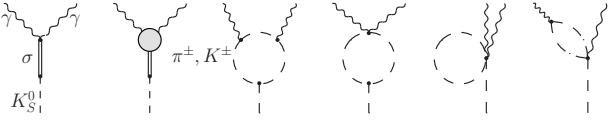}
\caption{\label{fig:loops} Leading order diagrams for 
$K_{S}^{0}\to\gamma\gamma$ in $\chi\mathrm{PT}_{\sigma}$, including
finite loop graphs \cite{DAm86}. 
The grey vertex contains $\pi^\pm,\,K^\pm$ 
loops as in the four $\chi$PT$_3$ diagrams to the right.  An analogous 
set of diagrams contributes to $\gamma\gamma\to\pi^{0}\pi^{0}$.}
\end{figure}
From $\gamma\gamma\rightarrow\pi^{0}\pi^{0}$ and 
$K_{S}^{0}\rightarrow\gamma\gamma$ (Fig.~\ref{fig:loops}), we estimate
\begin{equation}
|g^{}_{K_S\sigma}| \approx 4.4\times 10^{3}\,\mathrm{keV}^{2}
\label{gksig} \end{equation} 
to about 30\% precision and so, to the extent that $g_{\sigma NN}$
and hence $F_\sigma$ can be determined,
\begin{equation}
\left|A_{\sigma\textrm{-pole}}\right|
\approx 0.34\,\mathrm{keV}.
\end{equation}
This accounts for the large $I=0\ \pi\pi$ amplitude $A_0$ \cite{PDG}
\begin{equation}
|A_{0}|_{\mathrm{expt.}} = 0.33\,\mathrm{keV}
\end{equation}
compared with $A_2$. So we conclude that the observed ratio 
$|A_0/A_2| \simeq 22$ is mostly due to the dilaton-pole diagram 
of Fig.~\ref{fig:k_sig_pipi}, that $g_{8} = \sum_n g_{8n}$ and $g_{27}$ 
may have similar magnitudes as simple calculations indicate, and
that only $g_{27}$ can be fixed precisely (from 
$K^{+}\rightarrow\pi^{+}\pi^{0}$). 

Consequently, the leading order of $\chi$PT$_\sigma$ solves the 
$\Delta I = 1/2$ problem for kaon decays. The chiral Ward identities 
which relate the on-shell $K \to 2\pi$ and $K \to \pi$ amplitudes have 
extra terms due to $\sigma$ poles, but the no-tadpoles theorem \cite{RJC86}
is still valid:
\begin{equation}
\langle K | \mathcal{H}_{\mathrm{weak}} |\mathrm{vac}\rangle
= O\bigl(m_s^2 - m_d^2\bigr)\,, \ K\mbox{ on shell}.
\end{equation}

The presence of the $\sigma\gamma\gamma$ vertex in Fig.~\ref{fig:loops} leads 
us to apply the electromagnetic trace anomaly \cite{RJC72,Ell72}
\begin{gather}
\big.\theta^\mu_\mu\big|_{\mathrm{strong\,+\,e'mag}}
= \theta^{\mu}_{\mu} + (R\alpha/6\pi) F_{\mu\nu} F^{\mu\nu} , \notag \\
R=\left.\frac{\sigma(e^{+}e^{-}\rightarrow\mathrm{hadrons})}%
{\sigma(e^{+}e^{-}\rightarrow\mu^{+}\mu^{-})}\right|_\mathrm{high-energy}
\label{eqn:em_anomaly}
\end{gather}
to QCD at the infrared fixed point $\alpha_s = \alpha_\mathrm{IR}$. Here 
$F_{\mu\nu}$ and $\alpha$ are the electromagnetic field strength tensor 
and fine-structure constant.  

The $\chi$PT$_\sigma$ result for the $\sigma\gamma\gamma$ amplitude is affected 
by the observation \cite{DAm86} that in $\gamma\gamma$ channels, charged 
$\pi, K$ loop diagrams are finite in the chiral limit. This means that
they are of the \emph{same} order as $\sigma$-pole diagrams: partial 
conservation of the dilatation current is \emph{not} equivalent to simple 
$\sigma$-pole dominance in $\gamma\gamma$ channels. The electromagnetic 
trace anomaly remains the same (with $R \to R_\mathrm{IR}$), but because 
of the extra $(\pi^\pm, K^\pm)$ loops in 
$\langle\gamma\gamma|\theta^\mu_\mu|\mathrm{vac}\rangle$, the 
$\sigma\gamma\gamma$ coupling is proportional to $R_\mathrm{IR} - \frac{1}{2}$, 
not $R_\mathrm{IR}$:
\begin{equation}
\mathcal{L}_{\sigma\gamma\gamma} 
 = \bigl(R_{\mathrm{IR}} - \tfrac{1}{2}\bigr)\alpha(6\pi F_{\sigma})^{-1}
    \sigma F_{\mu\nu}F^{\mu\nu} \,. 
\label{Lgam}
\end{equation}

Evidently, data used to estimate $|g^{}_{K_S\sigma}|$ can also be used
to find a phenomenological value for $R_{\mathrm{IR}}$.  In a dispersive
analysis of $\gamma\gamma \to \pi\pi$, it was shown \cite{Pen06} that the 
residue of the $f_0(500)$ pole can be extracted from the Crystal Ball data 
\cite{Xal90}. We use an updated value \cite{Oll08} of the width 
$\Gamma(f_{0}\rightarrow\gamma\gamma) =  1.98^{+0.30}_{-0.24}$ keV. 
Within the large uncertainty due to that in $F_\sigma$, we find:
\begin{equation}
R_{\mathrm{IR}} \approx 5 \,.
\end{equation}
This result is a feature of the non-perturbative theory at 
$\alpha_\mathrm{IR}$, so it has \emph{nothing} 
to do with asymptotic freedom or the free-field formula $(N_f = 3)$
\begin{equation}
R(\alpha_s = 0) 
= \sum\{\mbox{quark charges}\}^2 = 2 \,.
\end{equation}

Notice that $\chi$PT$_\sigma$ relates amplitudes in the physical region
(\ref{phys}) to high-energy quantities like $\delta(\alpha_\mathrm{IR})$ 
and $R_{\mathrm{IR}}$ characteristic of \emph{massless} QCD at 
$\alpha_\mathrm{IR}$. Does QCD simplify in that limit and, unlike QED 
\cite{JBWA}, allow $\beta' \not= 0$ at the fixed point? 

Unfortunately, our analysis does not explain the
failure of chiral theory to account for non-leptonic $|\Delta S| = 1$
hyperon decays. We have shown that octet dominance is not necessary, 
but that makes no difference for hyperon decays: the Pati-Woo 
 $\Delta I = 1/2$ mechanism \cite{Pati71} forbids all
contributions from \textbf{27} operators.

\begin{acknowledgments}
We thank Ross Young and Peter Minkowski for valuable discussions at various 
stages of this work.  L.C.T.\ thanks Mary~K.~Gaillard and her colleagues for
their kind hospitality at the Berkeley Center for Theoretical Physics, 
where part of this work was completed.  L.C.T.\ is supported in part by 
the Australian-American Fulbright Commission, the Australian Research 
Council, the U.S. Department of Energy under Contract DE-AC02-05CH11231, 
and the National Science Foundation under grant PHY-0457315.
\end{acknowledgments}

\end{document}